\documentclass[preprint,aps,prb,showpacs,amsmath,amssymb,a4paper,superscriptaddress,11pt]{revtex4-1}
\usepackage{graphicx}			
\usepackage{bm}           
\usepackage{natbib}       
\usepackage{color}
\usepackage{braket}
\usepackage{amsmath}

\begin{document}

\title{{\LARGE A fault-tolerant addressable spin qubit in a natural silicon quantum dot}}

\author{K.~Takeda}
\affiliation{RIKEN, Center for Emergent Matter Science (CEMS),Wako-shi, Saitama, 351-0198, Japan}
\author{J.~Kamioka}
\affiliation{Department of Physical Electronics and Quantum Nanoelectronics Research Center, Tokyo Institute of Technology, O-okayama, Meguro-ku, Tokyo 152-8552, Japan}
\author{T.~Otsuka}
\affiliation{RIKEN, Center for Emergent Matter Science (CEMS),Wako-shi, Saitama, 351-0198, Japan}
\author{J.~Yoneda}
\affiliation{RIKEN, Center for Emergent Matter Science (CEMS),Wako-shi, Saitama, 351-0198, Japan}
\author{T.~Nakajima}
\affiliation{RIKEN, Center for Emergent Matter Science (CEMS),Wako-shi, Saitama, 351-0198, Japan}
\author{M.R.~Delbecq}
\affiliation{RIKEN, Center for Emergent Matter Science (CEMS),Wako-shi, Saitama, 351-0198, Japan}
\author{S.~Amaha}
\affiliation{RIKEN, Center for Emergent Matter Science (CEMS),Wako-shi, Saitama, 351-0198, Japan}
\author{G.~Allison}
\affiliation{RIKEN, Center for Emergent Matter Science (CEMS),Wako-shi, Saitama, 351-0198, Japan}
\author{T.~Kodera}
\affiliation{Department of Physical Electronics and Quantum Nanoelectronics Research Center, Tokyo Institute of Technology, O-okayama, Meguro-ku, Tokyo 152-8552, Japan}
\author{S.~Oda}
\affiliation{Department of Physical Electronics and Quantum Nanoelectronics Research Center, Tokyo Institute of Technology, O-okayama, Meguro-ku, Tokyo 152-8552, Japan}
\author{S.~Tarucha}
\affiliation{RIKEN, Center for Emergent Matter Science (CEMS),Wako-shi, Saitama, 351-0198, Japan}
\affiliation{Department of Applied Physics, The University of Tokyo, Hongo, Bunkyo-ku, Tokyo, 113-8656, Japan}

\date{\today}


\maketitle


{\bf Abstract}

Fault-tolerant quantum operation is a key requirement for the development of quantum computing. 
This has been realized in various solid-state systems including isotopically purified silicon which provides a nuclear spin free environment for the qubits, but not in industry standard natural (unpurified) silicon.
Here we demonstrate an addressable fault-tolerant qubit using a natural silicon double quantum dot with a micromagnet optimally designed for fast spin control. 
This optimized design allows us to achieve the optimum Rabi oscillation quality factor $Q=140$ at a Rabi frequency of 10 MHz in the frequency range two orders of magnitude higher than that achieved in previous studies. 
This leads to a qubit fidelity of 99.6 \%, which is the highest reported for natural silicon qubits and comparable to that obtained in isotopically purified silicon quantum-dot-based qubits. This result can inspire contributions from the industrial and quantum computing communities.
\\

Since the proposal of spin qubits using electrons confined in quantum dots \cite{Loss1998}, a great deal of effort has been made to implement quantum-dot-based spin qubits in a variety of semiconductors such as group III-V\cite{Petta2005, Koppens2006, Nowack2007, Pioro-Ladriere2008, Nadj-Perge2010, VandenBerg2013} or natural silicon \cite{Maune2012, Wu2014, Kawakami2014}. However, the quantum gate fidelities in these qubits are limited mainly due to the short coherence time ($T_2 ^* < 0.1$ $\mu$s \cite{Petta2005, Nadj-Perge2010, VandenBerg2013} for III-V semiconductors and $T_2 ^* <1$ $\mu$s for natural silicon\cite{Maune2012, Wu2014, Kawakami2014}) caused by the nuclear spin magnetic field fluctuations.
To obtain a qubit fidelity higher than the quantum error correction threshold for fault-tolerant quantum computing \cite{Knill2000, Fowler2009}, a straightforward approach is to prolong the qubit decay time $T^{\mathrm{Rabi}} _2$ or shorten the $\pi$ rotation time $T_{\pi}$ to increase the qubit Rabi oscillation quality factor $Q=T^{\mathrm{Rabi}} _2/T_{\pi}$ since it determines the upper bound of the qubit fidelity.
The first approach has been implemented in an isotopically purified silicon qubit with a long coherence time ($T_2 ^* \sim 120$ $\mu$s and $T_2 ^{\mathrm{Rabi}} \sim 380$ $\mu$s with 0.08 \% residual $^{29}$Si nuclear spins \cite{Veldhorst2014}), 
while leaving the spin control time ($T_{\pi}=1.6$ $\mu$s \cite{Veldhorst2014}) much slower than the other quantum-dot-based spin qubits ($T_{\pi} \sim 0.005$ $\mu$s \cite{VandenBerg2013, Yoneda2014}). 

For the realization of fault-tolerant qubits in more common materials such as natural silicon, one important issue that must be resolved is the slow spin control time \cite{Veldhorst2014, Veldhorst2015}. 
The key parameter to realize the fast spin control is a large (effective) oscillating magnetic field to drive the spin resonance, while keeping the applied microwave power small enough to suppress unwanted effects such as photon-assisted tunneling \cite{Koppens2006, VandenBerg2013} or heating.
However, because silicon does not have any strong spin driving mechanisms such as spin-orbit interaction\cite{VandenBerg2013, Laird2013}, an on-chip coplanar stripline is commonly used as a method to generate an oscillating magnetic field  \cite{Veldhorst2014, Veldhorst2015} despite being unsuitable for generating a large magnetic field. Alternatively, a micromagnet technique \cite{Tokura2006} can be used to implement a material-independent artificial strong spin-orbit coupling\cite{Yoneda2014} resulting in a much stronger effective magnetic field. 
Therefore the micromagnet technique may help to increase the quality factor for natural silicon quantum dot  qubits comparable to that in isotopically purified silicon but in a much high Rabi frequency range.
Recently, the technique has been applied to a silicon single quantum dot \cite{Kawakami2014} and 
an improvement of the Rabi frequency ($f_{\mathrm{Rabi}}$) by an order of magnitude was achieved, although the device structure was not appropriately optimized for fast and addressable control. 

Here we report an addressable fault-tolerant qubit using a natural silicon double quantum dot with a micromagnet optimally designed for fast spin control. 
From the microwave spectroscopy, the resonance frequency difference of about 800 MHz is obtained for the two electron spins confined in each quantum dot.
This result shows the good addressability of our qubits since the obtained frequency difference is about two orders of magnitude larger than our fast Rabi frequency and the crosstalk error is as small as 0.02 \% for our typical Rabi frequency of 10 MHz.
Next, the qubit dephasing time $T_2 ^*$ is measured by Ramsey interference. It shows a standard Gaussian decay with  $T_2 ^*$ of about 2 $\mu$s caused by the nuclear spin fluctuations.
The two-axis single-qubit control is confirmed by the observation of the shift of Ramsey fringe by the microwave burst phase modulation.
To operate the qubit much faster than its decay rate or to maximize the Rabi oscillation quality factor, the microwave amplitude ($A_{\mathrm{MW}}$) dependence of $f_{\mathrm{Rabi}}$ and $T^{\mathrm{Rabi}} _2$ is measured.
The optimized $A_{\mathrm{MW}}$ corresponds to $f_{\mathrm{Rabi}} \sim 10$ MHz, which is about two orders of magnitude faster than the reported value for an isotopically purified silicon quantum dot \cite{Veldhorst2014},  owing to the effectiveness of the magnetic field generation by our optimized micromagnet.
Finally, Clifford-based randomized benchmarking is performed to determine the qubit fidelity. At the optimized $A_{\mathrm{MW}}$, 
an average single-qubit fidelity of 99.6 \% is obtained. The qubit fidelity is reduced for the condition with shorter $T_{\pi}$ and smaller $Q$, which indicates that it is limited by heating at large $A_{\mathrm{MW}}$.\\

{\bf Results}

{\bf Quantum dot characterization.}
Our double quantum dot in silicon is formed by depleting a two-dimensional electron gas in an undoped natural Si/SiGe heterostructure by lithographically defined electrostatic gates (Fig. 1A). 
A 250 nm thick cobalt micromagnet is placed on top of the device to induce a stray magnetic field around the quantum dot. 
To achieve the fast and addressable control of single-electron spins using electric dipole spin resonance (EDSR), the micromagnet is designed to maximize the slanting magnetic field, $\mathrm{d}B^{\mathrm{MM}} _{\mathrm{y}} /\mathrm{d}z$, and the local Zeeman field difference between the two dots \cite{Yoneda2014}, $\Delta B_{\mathrm{z}} =  | B^{\mathrm{MM, R}}_{\mathrm{z}} - B^{\mathrm{MM, L}}_{\mathrm{z}}|$, where $B^{\mathrm{MM, R(L)}}$ denotes the stray magnetic field at the right (left) dot position. 
This micromagnet enables a slanting field several times larger compared to previous works \cite{Kawakami2014} (the micromagnet simulation is provided in Supplementary Section 1).
A nearby sensor quantum dot coupled to a radio frequency tank circuit allows rapid measurement of the double quantum dot charge configuration \cite{Reilly2007}. 
The sample is cooled down using a dilution refrigerator to a base electron temperature of 120 mK estimated from the line width of the dot transport. 
An in-plane external magnetic field $B_{\mathrm{ext}}$ is applied using a superconducting magnet. 
The double quantum dot is tuned to the (1,1) charge state where each dot hosts only one electron (Fig. 1C). 
Single-shot measurement of the spin state is performed using energy selective readout technique \cite{Elzerman2004} (Supplementary Fig. S1A). 

{\bf Qubit spectroscopy and Rabi oscillation.}
Figure 1B shows the pulse sequence for the spin control.
First, the spin-down state is initialized by applying gate voltages such that only the ground spin-down state can tunnel into the dot.
Next, the gate voltages are pulsed so that the electrons confined in the dot are pushed deep in Coulomb blockade.
Then, a microwave burst with a frequency of $f_{\mathrm{MW}}$ is applied to the gate C to induce EDSR.
Finally the gate voltages are pulsed back to the spin readout position where only a spin-up electron can tunnel out to the reservoir.
An additional emptying (or compensation) stage is used to keep the DC offset of the pulse to zero.
When the microwave burst is applied to the gate, the wave function 
of electrons confined in the dot oscillates spatially in the slanting magnetic field induced by the micromagnet, resulting in an effective oscillating magnetic field $B_{\mathrm{AC}}$ perpendicular to the static magnetic field $B_0 = B_{\mathrm{ext}} + B^{\mathrm{MM}}_{\mathrm{z}}$.
At the condition where $hf_{\mathrm{MW}} = g \mu B_0$, EDSR takes place. 
The resonance conditions are different for each dot by an amount proportional to $\Delta B_{\mathrm{z}}$, and therefore, 
the resonances of each dot can be addressed independently.

Figure 1D shows the spin-up probabilities for both right 
($P_{\uparrow} ^{\mathrm{R}}$, red signal) and left ($P_{\uparrow} ^{\mathrm{L}}$, blue signal) dot as a function of  $B_{\mathrm{ext}}$ and $f_{\mathrm{MW}}$.
For this measurement, a rectangular microwave burst with a fixed duration of $t_{\mathrm{p}} = 3$ $\mu$s is applied.
There are two clear resonance lines corresponding to each dot, which are separated by  $\Delta B _{\mathrm{z}} \sim 30$ mT or 800 MHz consistent with our micromagnet simulation (Supplementary Fig. S1C.).
The observed 800 MHz splitting is approximately two orders of magnitude larger than the value obtained for the spin-orbit mediated Stark shift of $g^*$-factor in a silicon quantum dot spin qubit without a micromagnet \cite{Veldhorst2014}.
For multiple qubit systems, 
it is crucial to have such a large frequency splitting to operate the qubit independently without crosstalk since the effect of the driving field decays with $(f_{\mathrm{Rabi}})^2   / ((\Delta f) ^2 + (f_{\mathrm{Rabi}})^2)$, where $\Delta f = f_{\mathrm{res}} ^{\mathrm{R(L)}} - f _{\mathrm{MW}}$ is the frequency detuning from the centre resonance frequency and $f_{\mathrm{Rabi}}$ is the Rabi frequency.
For our typical $f_{\mathrm{Rabi}}$ of 10 MHz, the 800 MHz splitting yields a crosstalk operation of the idle qubit with an amplitude as small as 0.02 \% of the operated qubit.
For the following measurements, we mainly focus on the left quantum dot as the sensitivity of our charge sensor is significantly higher due to the design of our device. 
However, the right quantum dot shows similar results, as detailed in the Supplementary Section 2.

The coherent evolution of the spin state is measured by changing the microwave duration at the resonance frequency (Fig. 1E).
The red triangles show the experimental data and the black solid line shows a fit with an exponentially damped sinusoidal function with a Rabi oscillation decay time $T_2 ^{\mathrm{Rabi}}$ (details on the fitting procedure are provided in Supplementary Section 3).
Note that the $T_2 ^{\mathrm{Rabi}}$ of a strongly driven qubit is different from the standard $T_2 ^*$ time measured from Ramsey interference since the influence of nuclear spin fluctuations is suppressed by the Rabi driving field \cite{Yoneda2014}.
When $f_{\mathrm{MW}}$ is detuned from the resonance frequency, 
the qubit rotates around a tilted axis in the Bloch sphere. This results in faster rotation at detuned $f_{\mathrm{MW}}$ and the chevron pattern shown in Fig. 1F.

{\bf Qubit coherence measurement.}
Next, characterisation of the qubit coherence time is performed using a Ramsey interference technique (Fig. 2A). 
First the spin is initialized in the down state. Then, a $\pi /2$ pulse is applied to rotate the spin to the equator of the Bloch sphere where it accumulates a phase error for the wait time $t_{\mathrm{w}}$. Finally the spin state is rotated by the second $\pi /2$ pulse to project the phase error to the $z$-axis.
Figure 2B shows a Ramsey measurement result which shows well-defined Ramsey fringes.
From the decrease of the Ramsey fringe amplitude as a function of $t_{\mathrm{w}}$, the value of the dephasing time $T_2 ^*$ can be obtained (Fig. 2C). 
The red triangles show measured data and the black solid line is a fit with a Gaussian decay function. 
The obtained $T_2 ^*$ for both dots are roughly 2 $\mu$s (the right dot measurement data is provided in Supplementary Section 1) and 
this is the longest value observed in isotopically natural silicon spin qubit systems \cite{Kawakami2014, Maune2012, Wu2014}, despite the fact that the large slanting magnetic field can make the spin state more sensitive to charge noise  \cite{Neumann2015, Kha2015}. 
Two-axis control in the Bloch sphere is required for arbitrary spin-qubit control.
This is demonstrated by modulating the phase $\varphi$ of the second microwave burst in the Ramsey measurement.
The observed shift of the Ramsey fringe in Fig. 2D corresponds to the change of rotation angle of the second $\pi /2$ rotation.

{\bf Microwave power dependence.}
To improve the qubit fidelity by suppressing the influence of dephasing, it is straightforward to maximize the Rabi oscillation quality factor $Q=T ^{\mathrm{Rabi}} _2 / T_{\pi}$ by applying a larger microwave excitation to decrease the $\pi$ rotation time $T_{\pi}$. 
However, it has been reported that too large a microwave excitation can cause a significant dephasing 
due to photon-assisted tunnelling \cite{Yoneda2014, Koppens2005, Nowack2007}, therefore it is important to characterize the microwave amplitude ($A_{\mathrm{MW}}$) dependence of the Rabi oscillation. 
As $A_{\mathrm{MW}}$ is increased, the oscillation period becomes shorter, however, the signal is damped more rapidly (Fig. 3A). 
By fitting the data of the damped oscillations with exponentially decaying functions (see Supplementary Section 3 for the details), 
$f_{\mathrm{Rabi}}$ and $T _2 ^{\mathrm{Rabi}}$ are extracted as a function of $A_{\mathrm{MW}}$ (Fig. 3B). 
$f_{\mathrm{Rabi}}$ increases with $A_{\mathrm{MW}}$ linearly when $A_{\mathrm{MW}}$ is smaller than about 0.3, but finally shows a saturation \cite{Yoneda2014}. The obtained maximum $f_{\mathrm{Rabi}}$ is about 35 MHz, showing an improvement of one or two orders of magnitude from previous experiments \cite{Kawakami2014, Veldhorst2014}.

$T _2 ^{\mathrm{Rabi}}$ shows a significant decrease when $A_{\mathrm{MW}}$ becomes larger. Here the photon assisted tunnelling mechanism may be ruled out since the decrease of $T _2 ^{\mathrm{Rabi}}$ does not depend strongly on the depth of Coulomb blockade or the operation point (Supplementary Fig. S5). Alternatively, heating due to the microwave burst, which causes the reduction of $T_2$ rather than $T_2^*$, can be a dominant source of the observed decay as previously observed in a singlet-triplet qubit \cite{Dial2013}. 
Figure 3D shows the quality factor $Q$ of the Rabi oscillations as a function of $A_{\mathrm{MW}}$. 
From this data, the optimal working point for the qubit operation is estimated to be $A_{\mathrm{MW}} \sim 0.2$  where $f_{\mathrm{Rabi}} \sim 10$ MHz and $Q \sim 140$ are obtained.
The obtained maximum $Q$ is in the same range as the one in an isotopically purified silicon quantum dot with two orders of magnitude slower $f_{\mathrm{Rabi}}$ and two orders of magnitude longer $T_2 ^{\mathrm{Rabi}}$ \, \cite{Veldhorst2014}. 

{\bf Qubit fidelity measurement by randomized benchmarking.}
Finally, the single-qubit control fidelity is characterized via randomized benchmarking \cite{Knill2008} using 
the measurement sequence shown in Fig. 4A. The reference sequence includes $m$ random Clifford gates and one recovery Clifford gate chosen such that the ideal final spin state becomes an eigenstate of $\sigma _{\mathrm{z}}$. 
The interleaving sequence \cite{Magesan2012a} is used to determine the fidelity of each single-step Clifford gate $C_{\mathrm{test}}$.
The pulse envelope is shaped to a Gaussian (truncated at $\pm 2 \sigma$) to minimize its spectral width for suppressing pulse errors \cite{Bauer1984}. 
An interval of 6 ns between gate operations is used to avoid pulse overlap. 
Fig 4B shows the reference randomized benchmarking measurement to determine the averaged Clifford gate fidelity. 
By fitting the reference measurement with an exponentially decaying curve $F(m)=A(2F_{\mathrm{c}}-1)^m$, where $A$ is the visibility  and $F_{\mathrm{c}}$ is the Clifford gate fidelity per step which corresponds to 1.875 single Clifford gates (the details are provided in Supplementary Section 5).
At the optimized $A_{\mathrm{MW}}$ to maximize $Q$ (the red point in the inset of Fig. 4B), we obtain  $F_{\mathrm{c}} ^{\mathrm{single}} = 99.6$ \% which is 
above the threshold for fault-tolerant quantum computing \cite{Fowler2009} and comparable to the value reported in an isotopically purified silicon quantum dot \cite{Veldhorst2014}.
Note $F_{\mathrm{c}} ^{\mathrm{single}}$ is decreased for a large $A_{\mathrm{MW}}$ or shorter $T_{\pi}$ (the blue point in the inset of Fig. 4B), indicating $Q$ is a good indicator to optimize $A_{\mathrm{MW}}$ and $F_{\mathrm{c}} ^{\mathrm{single}}$. 
Fig 4C shows interleaving measurements for the single-step Gaussian Clifford gates.
As expected from the microwave heating, it is found that the fidelities for $\pi /2$ gates are higher than those of $\pi$ gates.
This result is consistent with our qubit decay time limiting mechanism, the microwave effect.\\

{\bf Discussion}

The Si/SiGe double quantum dot used in this work contains an optimized micromagnet that enables Rabi frequencies nearly two orders of magnitude faster than the reported values in previous works without causing too much decoherence caused by unwanted photon-assisted tunneling or heating effects at high microwave amplitudes.
From the microwave amplitude dependence measurement, we find the optimum working point to maximize the Rabi oscillation quality factor $Q$. 
The maximum $Q$ of $140$ is obtained at $f_{\mathrm{Rabi}}= 10$ MHz, which is much faster than the qubit decay rate ($1/T_2 ^{\mathrm{Rabi}} \sim 140$ kHz). 
Together with the large qubit resonance frequency difference due to the large micromagnet inhomogeneous field ($\Delta f_{\mathrm{res}} \sim 800$ MHz), we can implement fault-tolerant, fast and addressable single-spin qubit operations even without the use of rare isotopically purified silicon.
This may facilitate the realization of a large scale quantum processor using existing industry standard silicon nano-fabrication techniques.
Our micromagnet technique can also be applied to isotopically purified silicon to further enhance the Rabi oscillation quality factor $Q$ and the qubit fidelity. We also note that it is possible to realize further  enhancements by several trivial optimizations with natural silicon devices ({\it e.g.} decreasing the distance between the dot and the micromagnet \cite{Yoneda2014} or changing the gate geometry to increase the microwave gate lever arm) to increase the effective AC magnetic field strength and $f_{\mathrm{Rabi}}$.
Although the effect of charge noise induced dephasing is not clearly observed in this work, it is important to balance the effect of the Rabi frequency enhancement and the dephasing caused by the charge noise to maximize $Q$ and the resulting qubit fidelity.
\bibliography{library}

\begin{thebibliography}{10}
\expandafter\ifx\csname url\endcsname\relax
  \def\url#1{\texttt{#1}}\fi
\expandafter\ifx\csname urlprefix\endcsname\relax\def\urlprefix{URL }\fi
\providecommand{\bibinfo}[2]{#2}
\providecommand{\eprint}[2][]{\url{#2}}

\bibitem{Loss1998}
\bibinfo{author}{Loss, D.} \& \bibinfo{author}{DiVincenzo, D.~P.}
\newblock \bibinfo{title}{{Quantum computation with quantum dots}}.
\newblock \emph{\bibinfo{journal}{Physical Review A}}
  \textbf{\bibinfo{volume}{57}}, \bibinfo{pages}{120--126}
  (\bibinfo{year}{1998}).

\bibitem{Petta2005}
\bibinfo{author}{Petta, J.~R.} \emph{et~al.}
\newblock \bibinfo{title}{{Coherent manipulation of coupled electron spins in
  semiconductor quantum dots.}}
\newblock \emph{\bibinfo{journal}{Science (New York, N.Y.)}}
  \textbf{\bibinfo{volume}{309}}, \bibinfo{pages}{2180--4}
  (\bibinfo{year}{2005}).

\bibitem{Koppens2006}
\bibinfo{author}{Koppens, F. H.~L.} \emph{et~al.}
\newblock \bibinfo{title}{{Driven coherent oscillations of a single electron
  spin in a quantum dot.}}
\newblock \emph{\bibinfo{journal}{Nature}} \textbf{\bibinfo{volume}{442}},
  \bibinfo{pages}{766--71} (\bibinfo{year}{2006}).

\bibitem{Nowack2007}
\bibinfo{author}{Nowack, K.~C.}, \bibinfo{author}{Koppens, F. H.~L.},
  \bibinfo{author}{Nazarov, Y.~V.} \& \bibinfo{author}{Vandersypen, L. M.~K.}
\newblock \bibinfo{title}{{Coherent control of a single electron spin with
  electric fields.}}
\newblock \emph{\bibinfo{journal}{Science (New York, N.Y.)}}
  \textbf{\bibinfo{volume}{318}}, \bibinfo{pages}{1430--3}
  (\bibinfo{year}{2007}).

\bibitem{Pioro-Ladriere2008}
\bibinfo{author}{Pioro-Ladri{\`{e}}re, M.} \emph{et~al.}
\newblock \bibinfo{title}{{Electrically driven single-electron spin resonance
  in a slanting Zeeman field}}.
\newblock \emph{\bibinfo{journal}{Nature Physics}}
  \textbf{\bibinfo{volume}{4}}, \bibinfo{pages}{776--779}
  (\bibinfo{year}{2008}).

\bibitem{Nadj-Perge2010}
\bibinfo{author}{Nadj-Perge, S.}, \bibinfo{author}{Frolov, S.~M.},
  \bibinfo{author}{Bakkers, E. P. A.~M.} \& \bibinfo{author}{Kouwenhoven,
  L.~P.}
\newblock \bibinfo{title}{{Spin-orbit qubit in a semiconductor nanowire.}}
\newblock \emph{\bibinfo{journal}{Nature}} \textbf{\bibinfo{volume}{468}},
  \bibinfo{pages}{1084--7} (\bibinfo{year}{2010}).

\bibitem{VandenBerg2013}
\bibinfo{author}{van~den Berg, J.} \emph{et~al.}
\newblock \bibinfo{title}{{Fast Spin-Orbit Qubit in an Indium Antimonide
  Nanowire}}.
\newblock \emph{\bibinfo{journal}{Physical Review Letters}}
  \textbf{\bibinfo{volume}{110}}, \bibinfo{pages}{066806}
  (\bibinfo{year}{2013}).

\bibitem{Maune2012}
\bibinfo{author}{Maune, B.~M.} \emph{et~al.}
\newblock \bibinfo{title}{{Coherent singlet-triplet oscillations in a
  silicon-based double quantum dot.}}
\newblock \emph{\bibinfo{journal}{Nature}} \textbf{\bibinfo{volume}{481}},
  \bibinfo{pages}{344--7} (\bibinfo{year}{2012}).

\bibitem{Wu2014}
\bibinfo{author}{Wu, X.} \emph{et~al.}
\newblock \bibinfo{title}{{Two-axis control of a singlet-triplet qubit with an
  integrated micromagnet.}}
\newblock \emph{\bibinfo{journal}{Proceedings of the National Academy of
  Sciences of the United States of America}} \textbf{\bibinfo{volume}{111}},
  \bibinfo{pages}{11938--42} (\bibinfo{year}{2014}).

\bibitem{Kawakami2014}
\bibinfo{author}{Kawakami, E.} \emph{et~al.}
\newblock \bibinfo{title}{{Electrical control of a long-lived spin qubit in a
  Si/SiGe quantum dot.}}
\newblock \emph{\bibinfo{journal}{Nature nanotechnology}}
  \textbf{\bibinfo{volume}{9}}, \bibinfo{pages}{666--670}
  (\bibinfo{year}{2014}).

\bibitem{Knill2000}
\bibinfo{author}{Knill, E.}, \bibinfo{author}{Laflamme, R.},
  \bibinfo{author}{Martinez, R.} \& \bibinfo{author}{Tseng, C.}
\newblock \bibinfo{title}{{An algorithmic benchmark for quantum information
  processing}}.
\newblock \emph{\bibinfo{journal}{Nature}} \textbf{\bibinfo{volume}{404}},
  \bibinfo{pages}{21--23} (\bibinfo{year}{2000}).

\bibitem{Fowler2009}
\bibinfo{author}{Fowler, A.~G.}, \bibinfo{author}{Stephens, A.~M.} \&
  \bibinfo{author}{Groszkowski, P.}
\newblock \bibinfo{title}{{High-threshold universal quantum computation on the
  surface code}}.
\newblock \emph{\bibinfo{journal}{Physical Review A}}
  \textbf{\bibinfo{volume}{80}}, \bibinfo{pages}{052312}
  (\bibinfo{year}{2009}).

\bibitem{Veldhorst2014}
\bibinfo{author}{Veldhorst, M.} \emph{et~al.}
\newblock \bibinfo{title}{{An addressable quantum dot qubit with fault-tolerant
  control-fidelity.}}
\newblock \emph{\bibinfo{journal}{Nature nanotechnology}} \bibinfo{pages}{1--5}
  (\bibinfo{year}{2014}).

\bibitem{Yoneda2014}
\bibinfo{author}{Yoneda, J.} \emph{et~al.}
\newblock \bibinfo{title}{{Fast Electrical Control of Single Electron Spins in
  Quantum Dots with Vanishing Influence from Nuclear Spins}}.
\newblock \emph{\bibinfo{journal}{Physical Review Letters}}
  \textbf{\bibinfo{volume}{113}}, \bibinfo{pages}{267601}
  (\bibinfo{year}{2014}).

\bibitem{Veldhorst2015}
\bibinfo{author}{Veldhorst, M.} \emph{et~al.}
\newblock \bibinfo{title}{{A two-qubit logic gate in silicon}}.
\newblock \emph{\bibinfo{journal}{Nature}} \textbf{\bibinfo{volume}{526}},
  \bibinfo{pages}{410--414} (\bibinfo{year}{2015}).

\bibitem{Laird2013}
\bibinfo{author}{Laird, E.~A.}, \bibinfo{author}{Pei, F.} \&
  \bibinfo{author}{Kouwenhoven, L.~P.}
\newblock \bibinfo{title}{{A valley-spin qubit in a carbon nanotube.}}
\newblock \emph{\bibinfo{journal}{Nature nanotechnology}}
  \textbf{\bibinfo{volume}{8}}, \bibinfo{pages}{565--8} (\bibinfo{year}{2013}).

\bibitem{Tokura2006}
\bibinfo{author}{Tokura, Y.}, \bibinfo{author}{van~der Wiel, W.~G.},
  \bibinfo{author}{Obata, T.} \& \bibinfo{author}{Tarucha, S.}
\newblock \bibinfo{title}{{Coherent Single Electron Spin Control in a Slanting
  Zeeman Field}}.
\newblock \emph{\bibinfo{journal}{Physical Review Letters}}
  \textbf{\bibinfo{volume}{96}}, \bibinfo{pages}{047202}
  (\bibinfo{year}{2006}).

\bibitem{Reilly2007}
\bibinfo{author}{Reilly, D.~J.}, \bibinfo{author}{Marcus, C.~M.},
  \bibinfo{author}{Hanson, M.~P.} \& \bibinfo{author}{Gossard, A.~C.}
\newblock \bibinfo{title}{{Fast single-charge sensing with a rf quantum point
  contact}}.
\newblock \emph{\bibinfo{journal}{Applied Physics Letters}}
  \textbf{\bibinfo{volume}{91}}, \bibinfo{pages}{162101}
  (\bibinfo{year}{2007}).

\bibitem{Elzerman2004}
\bibinfo{author}{Elzerman, J.~M.} \emph{et~al.}
\newblock \bibinfo{title}{{Single-shot read-out of an individual electron spin
  in a quantum dot}}.
\newblock \emph{\bibinfo{journal}{Nature}} \textbf{\bibinfo{volume}{430}},
  \bibinfo{pages}{431--435} (\bibinfo{year}{2004}).

\bibitem{Neumann2015}
\bibinfo{author}{Neumann, R.} \& \bibinfo{author}{Schreiber, L.~R.}
\newblock \bibinfo{title}{{Simulation of micro-magnet stray-field dynamics for
  spin qubit manipulation}}.
\newblock \emph{\bibinfo{journal}{Journal of Applied Physics}}
  \textbf{\bibinfo{volume}{117}}, \bibinfo{pages}{193903}
  (\bibinfo{year}{2015}).

\bibitem{Koppens2005}
\bibinfo{author}{Koppens, F. H.~L.} \emph{et~al.}
\newblock \bibinfo{title}{{Control and detection of singlet-triplet mixing in a
  random nuclear field.}}
\newblock \emph{\bibinfo{journal}{Science (New York, N.Y.)}}
  \textbf{\bibinfo{volume}{309}}, \bibinfo{pages}{1346--50}
  (\bibinfo{year}{2005}).

\bibitem{Dial2013}
\bibinfo{author}{Dial, O.~E.} \emph{et~al.}
\newblock \bibinfo{title}{{Charge Noise Spectroscopy Using Coherent Exchange
  Oscillations in a Singlet-Triplet Qubit}}.
\newblock \emph{\bibinfo{journal}{Physical Review Letters}}
  \textbf{\bibinfo{volume}{110}}, \bibinfo{pages}{146804}
  (\bibinfo{year}{2013}).

\bibitem{Knill2008}
\bibinfo{author}{Knill, E.} \emph{et~al.}
\newblock \bibinfo{title}{{Randomized benchmarking of quantum gates}}.
\newblock \emph{\bibinfo{journal}{Physical Review A}}
  \textbf{\bibinfo{volume}{77}}, \bibinfo{pages}{012307}
  (\bibinfo{year}{2008}).

\bibitem{Magesan2012a}
\bibinfo{author}{Magesan, E.} \emph{et~al.}
\newblock \bibinfo{title}{{Efficient Measurement of Quantum Gate Error by
  Interleaved Randomized Benchmarking}}.
\newblock \emph{\bibinfo{journal}{Physical Review Letters}}
  \textbf{\bibinfo{volume}{109}}, \bibinfo{pages}{080505}
  (\bibinfo{year}{2012}).

\bibitem{Bauer1984}
\bibinfo{author}{Bauer, C.}, \bibinfo{author}{Freeman, R.},
  \bibinfo{author}{Frenkiel, T.}, \bibinfo{author}{Keeler, J.} \&
  \bibinfo{author}{Shaka, A.~J.}
\newblock \bibinfo{title}{{Gaussian pulses}}.
\newblock \emph{\bibinfo{journal}{Journal of Magnetic Resonance}}
  \textbf{\bibinfo{volume}{58}}, \bibinfo{pages}{442--457}
  (\bibinfo{year}{1984}).

\end{thebibliography}


\begin{thebibliography}{1}
\expandafter\ifx\csname url\endcsname\relax
  \def\url#1{\texttt{#1}}\fi
\expandafter\ifx\csname urlprefix\endcsname\relax\def\urlprefix{URL }\fi
\providecommand{\bibinfo}[2]{#2}
\providecommand{\eprint}[2][]{\url{#2}}

\bibitem{Yoneda2014}
\bibinfo{author}{Yoneda, J.} \emph{et~al.}
\newblock \bibinfo{title}{{Fast Electrical Control of Single Electron Spins in
  Quantum Dots with Vanishing Influence from Nuclear Spins}}.
\newblock \emph{\bibinfo{journal}{Physical Review Letters}}
  \textbf{\bibinfo{volume}{113}}, \bibinfo{pages}{267601}
  (\bibinfo{year}{2014}).

\bibitem{Radia}
\bibinfo{title}{{We used MATHEMATICA RADIA package available at;
  http://www.esrf.fr/.}}

\bibitem{Yoneda2015}
\bibinfo{author}{Yoneda, J.} \emph{et~al.}
\newblock \bibinfo{title}{{Robust micromagnet design for fast electrical
  manipulations of single spins in quantum dots}}.
\newblock \emph{\bibinfo{journal}{Applied Physics Express}}
  \textbf{\bibinfo{volume}{8}} (\bibinfo{year}{2015}).

\bibitem{Dial2013}
\bibinfo{author}{Dial, O.~E.} \emph{et~al.}
\newblock \bibinfo{title}{{Charge Noise Spectroscopy Using Coherent Exchange
  Oscillations in a Singlet-Triplet Qubit}}.
\newblock \emph{\bibinfo{journal}{Physical Review Letters}}
  \textbf{\bibinfo{volume}{110}}, \bibinfo{pages}{146804}
  (\bibinfo{year}{2013}).

\bibitem{Muhonen2015}
\bibinfo{author}{Muhonen, J.~T.} \emph{et~al.}
\newblock \bibinfo{title}{{Quantifying the quantum gate fidelity of single-atom
  spin qubits in silicon by randomized benchmarking.}}
\newblock \emph{\bibinfo{journal}{Journal of physics. Condensed matter : an
  Institute of Physics journal}} \textbf{\bibinfo{volume}{27}},
  \bibinfo{pages}{154205} (\bibinfo{year}{2015}).

\bibitem{Magesan2012}
\bibinfo{author}{Magesan, E.}, \bibinfo{author}{Gambetta, J.~M.} \&
  \bibinfo{author}{Emerson, J.}
\newblock \bibinfo{title}{{Characterizing quantum gates via randomized
  benchmarking}}.
\newblock \emph{\bibinfo{journal}{Physical Review A}}
  \textbf{\bibinfo{volume}{85}}, \bibinfo{pages}{042311}
  (\bibinfo{year}{2012}).

\bibitem{Magesan2012a}
\bibinfo{author}{Magesan, E.} \emph{et~al.}
\newblock \bibinfo{title}{{Efficient Measurement of Quantum Gate Error by
  Interleaved Randomized Benchmarking}}.
\newblock \emph{\bibinfo{journal}{Physical Review Letters}}
  \textbf{\bibinfo{volume}{109}}, \bibinfo{pages}{080505}
  (\bibinfo{year}{2012}).

\end{thebibliography}

\

{\bf Acknowledgements}

We thank R. Sugawara and T. Obata for technical contributions. 
This work was supported financially by the Funding Program for World-Leading Innovative R \& D on Science and Technology (FIRST) from the Japan Society for the Promotion of Science, ImPACT Program of Council for Science, Toyota Physical \& Chemical Research Institute Scholars, RIKEN Incentive Research Project, the Grant-in-Aid for Research Young Scientists B, Yazaki Memorial Foundation for Science and Technology Research Grant, Japan Prize Foundation Research Grant, Advanced Technology Institute Research Grant, the Murata Science Foundation Research Grant and Kakenhi Grants in-Aid (Nos. 26709023 and 26630151).

\

{\bf Author contributions}

K.T. and J.K. fabricated the sample and performed the measurement.
K.T. analysed the data and wrote the manuscript with inputs from all of the other authors.
T.O., J.Y., T.N., M.R.D., S.A., G.A., T.K. and S.O. contributed to the sample fabrication, the measurement and the data analysis.
S.T. supervised the project.

\

{\bf Additional information}

Correspondence and requests for materials should be addressed to K.T. (kenta.takeda@riken.jp) or S.T. (tarucha@ap.t.u-tokyo.ac.jp).

\

{\bf Competing financial interests}

The authors declare no competing financial interests.

\newpage

\begin{figure*}[t]
\centering
\includegraphics*[width=1.0\columnwidth]{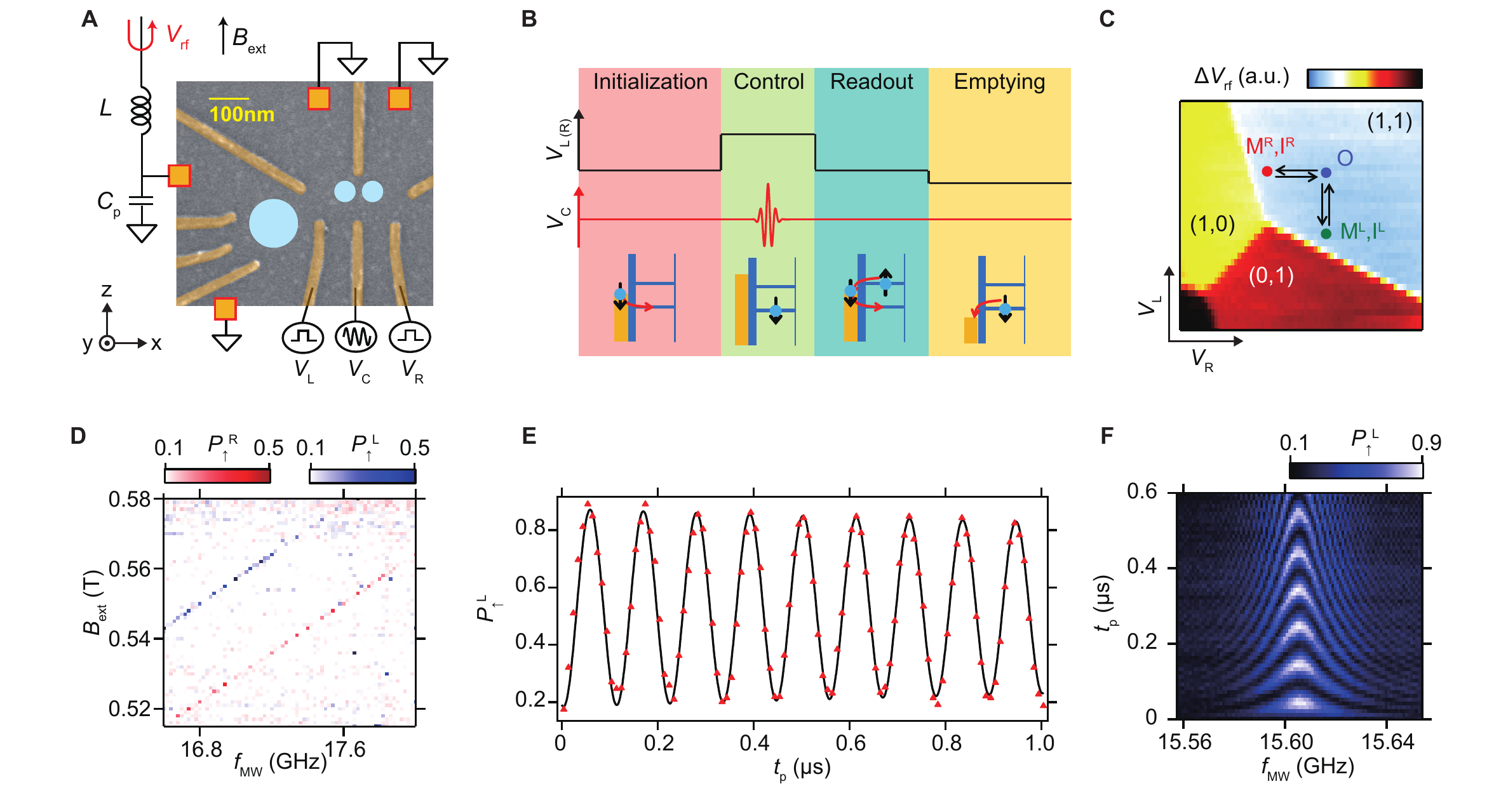}
\caption{
\label{Fig1} 
{\bf Device structure and EDSR measurement result}. 
{\bf A}, False color scanning electron micrograph of the device. The orange boxes represent ohmic contacts that are grounded during the measurements except for the one connected to the resonance circuit. Three of the gate electrodes (R, L and C) are connected to impedance matched high-frequency lines with cryogenic bias-tees.
{\bf B}, 
Schematic of the pulse sequence used for the EDSR measurement. The pulse sequence consists of 4 stages of initialization, control, readout and emptying.
{\bf C}, 
Charge stability diagram in the vicinity of the (1,1) charge configuration. $\mathrm{M^R, I^R}$ ($\mathrm{M^L, I^L}$) denote the measurement and initialization points for the right(left) quantum dot. O denotes the  operation point which is common for both right and left quantum dots. 
{\bf D}, Measurement of the EDSR signal as a function of $f_{\mathrm{MW}}$ and $B_{\mathrm{ext}}$. The blue line corresponds to the left dot resonance condition $hf_{\mathrm{MW}} = g\mu (B_{\mathrm{ext}} + B^{\mathrm{MM, L}}_{\mathrm{z}})$ and the red line corresponds to the right dot resonance condition $hf_{\mathrm{MW}} = g\mu (B_{\mathrm{ext}} + B^{\mathrm{MM, R}}_{\mathrm{z}})$.
{\bf E}, Rabi oscillation with $T_2 ^{\mathrm{Rabi}} \sim 8$  $\mu$s and $f _{\mathrm{Rabi}} \sim 9$ MHz measured  at $B_{\mathrm{ext}} = 0.505 $T and $f_{\mathrm{MW}}=15.6055$ GHz.
The red triangles show measurement data and the black solid line shows the fitting with an exponentially damped sine curve, $P_{\uparrow} ^{\mathrm{L}} (t_{\mathrm{p}}) = A \exp{(-t_{\mathrm{p}}/T_2 ^{\mathrm{Rabi}})} \sin {(2\pi f _{\mathrm{Rabi}} t_{\mathrm{p}} + \phi )} + B$ with $A$, $B$ and $T_2 ^{\mathrm{Rabi}}$ as fitting parameters.
{\bf F}, Measurement result of detuned Rabi oscillations which shows a typical chevron pattern.}
\end{figure*}

\begin{figure*}[!t]
\centering
\includegraphics*[width=0.5\columnwidth]{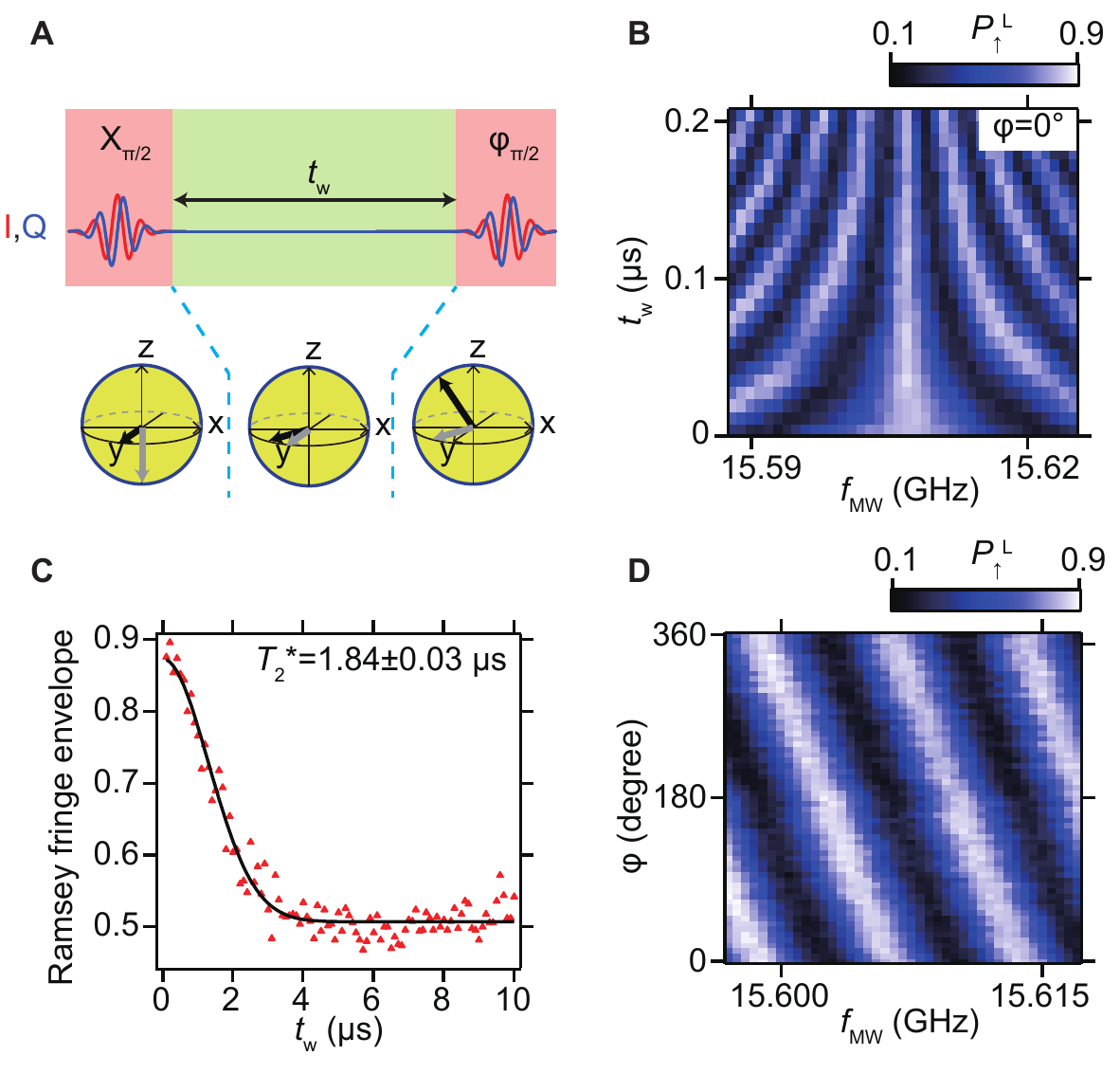}
\caption{\label{Fig2} 
{\bf Ramsey interference measurements}. 
{\bf A}, Schematic of the Ramsey measurement sequence. $\varphi$ denotes the phase of the second microwave burst relative to the first X$_{\pi /2}$ rotation. A rectangle or Gaussian microwave burst is applied to the gate C.
{\bf B}, Ramsey fringes measurement result. $B_{\mathrm{ext}}$ is fixed at 0.505 T. $\varphi$ is the phase of the second microwave burst relative to the first microwave burst.
{\bf C}, Ramsey fringes decay envelope extracted by sweeping $f_{\mathrm{MW}}$ at each fixed $t_{\mathrm{w}}$. The black solid line is a fit with a Gaussian decay function $P_{\uparrow} ^{\mathrm{L}} (t_{\mathrm{w}}) = A \exp{(-(t_{\mathrm{w}} /T_2 ^{*})^2)} + B$, where $A$ and $B$ are constants to account for the measurement and initialization fidelities.
{\bf D}, Demonstration of $\pi /2$ pulse around an arbitrary rotation axis in the xy-plane of the Bloch sphere.}
\end{figure*}

\begin{figure*}[!t]
\centering
\includegraphics*[width=0.5\columnwidth]{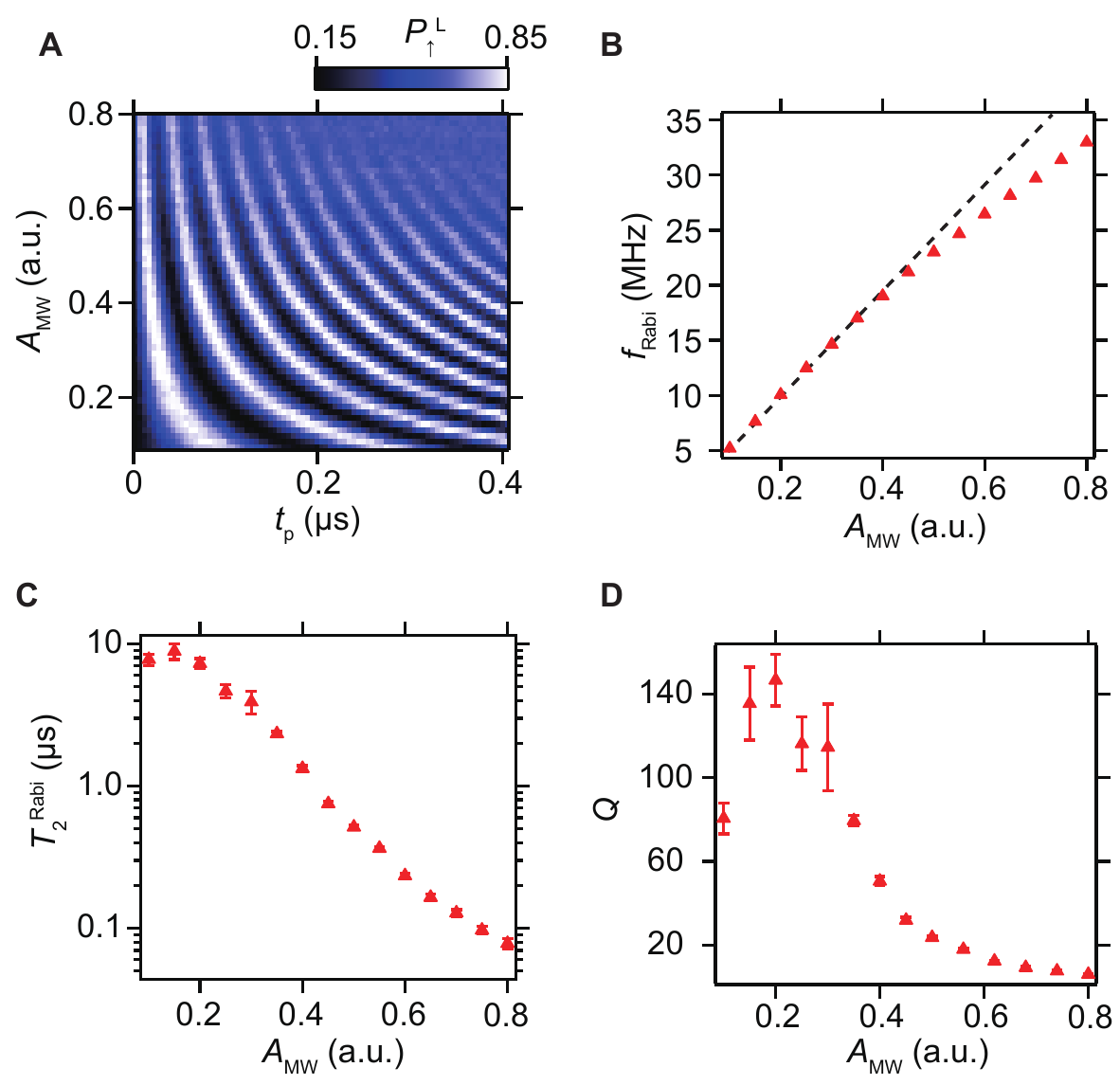}
\caption{\label{Fig3} 
{\bf Rabi oscillation power dependence}. 
{\bf A}, Microwave amplitude dependence of Rabi oscillations measured at $B_{\mathrm{ext}} = 0.505$ T and $f_{\mathrm{MW}} = 15.6055$ GHz. 
{\bf B}, Microwave amplitude dependence of the Rabi frequency $f_{\mathrm{Rabi}}$. The red triangles show the measured data and the black dotted line shows a linear fitting for the small amplitude data ($0.1 \leq A_{\mathrm{MW}} \leq 0.25$). The fitting error is smaller than the size of the symbols.
{\bf C}, Microwave amplitude dependence of the Rabi decay time $T_2 ^{\mathrm{Rabi}}$. 
Because the total evolution time of the data used for the fitting is relatively short ($t_p = 3$ $\mu$s), it shows large errors for small $A_{\mathrm{MW}}$ points. 
{\bf D}, Microwave amplitude dependence of the quality factor $Q=T_2 ^{\mathrm{Rabi}}/ T_{\pi}$. The error mainly comes from the uncertainty of $T_2 ^{\mathrm{Rabi}}$.
}
\end{figure*}

\begin{figure*}[!t]
\centering
\includegraphics*[width=0.5\columnwidth]{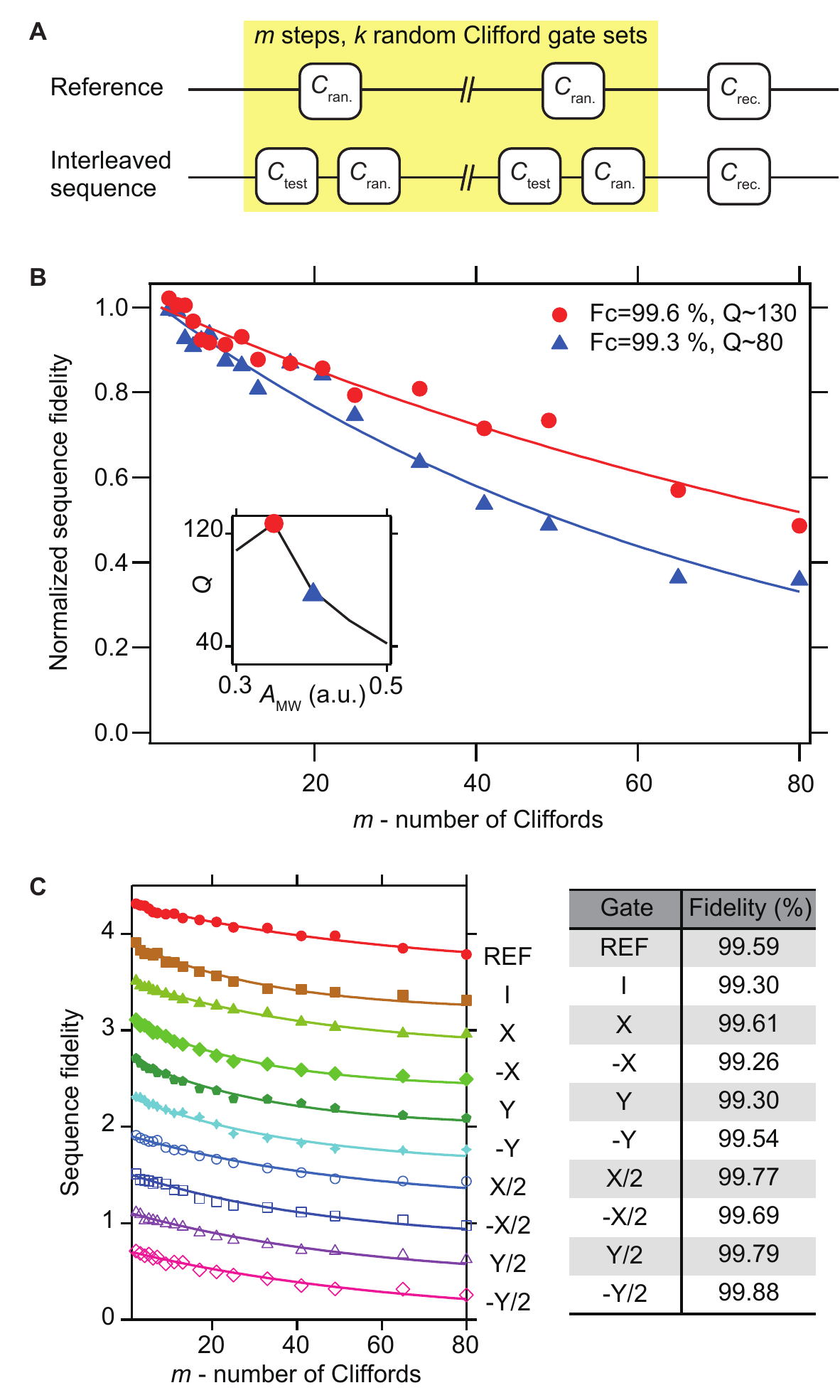}
\caption{\label{Fig4} 
{\bf Randomized benchmarking measurement}. 
{\bf A}, Schematic of the randomized benchmarking sequence. The upper is a reference sequence consisting of $m$ random Clifford gates. The lower is the interleaved sequence used to measure fidelities of a specific test Clifford gate $C_{\mathrm{test}}$. The sequence is repeated for $k=16$ choices of sequences to obtain one point.
{\bf B}, Reference randomized benchmarking for two different microwave amplitudes. The inset figure shows the quality factor measurement for Gaussian microwave burst, which shows quite similar result to the one for rectangle microwave burst.
{\bf C}, Interleaved randomized benchmarking for single-step Clifford gates. The right table shows fidelity measurement result for several single qubit gates. The fitting error of each gate fidelity is smaller than 0.1 \% for the reference and all interleaving measurements.}
\end{figure*}

\end{document}


\title{{\LARGE Supplementary information for a fault-tolerant addressable spin qubit in a natural silicon quantum dot}}

\author{K.~Takeda}
\affiliation{RIKEN, Center for Emergent Matter Science (CEMS),Wako-shi, Saitama, 351-0198, Japan}
\author{J.~Kamioka}
\affiliation{Quantum Nanoelectronics Research Center, Tokyo Institute of Technology, O-okayama, Meguro-ku, Tokyo 152-8552, Japan}
\author{T.~Otsuka}
\affiliation{RIKEN, Center for Emergent Matter Science (CEMS),Wako-shi, Saitama, 351-0198, Japan}
\author{J.~Yoneda}
\affiliation{RIKEN, Center for Emergent Matter Science (CEMS),Wako-shi, Saitama, 351-0198, Japan}
\author{T.~Nakajima}
\affiliation{RIKEN, Center for Emergent Matter Science (CEMS),Wako-shi, Saitama, 351-0198, Japan}
\author{M.R.~Delbecq}
\affiliation{RIKEN, Center for Emergent Matter Science (CEMS),Wako-shi, Saitama, 351-0198, Japan}
\author{S.~Amaha}
\affiliation{RIKEN, Center for Emergent Matter Science (CEMS),Wako-shi, Saitama, 351-0198, Japan}
\author{G.~Allison}
\affiliation{RIKEN, Center for Emergent Matter Science (CEMS),Wako-shi, Saitama, 351-0198, Japan}
\author{T.~Kodera}
\affiliation{Quantum Nanoelectronics Research Center, Tokyo Institute of Technology, O-okayama, Meguro-ku, Tokyo 152-8552, Japan}
\author{S.~Oda}
\affiliation{Quantum Nanoelectronics Research Center, Tokyo Institute of Technology, O-okayama, Meguro-ku, Tokyo 152-8552, Japan}
\author{S.~Tarucha}
\affiliation{RIKEN, Center for Emergent Matter Science (CEMS),Wako-shi, Saitama, 351-0198, Japan}
\affiliation{Department of Applied Physics, The University of Tokyo, Hongo, Bunkyo-ku, Tokyo, 113-8656, Japan}

\date{\today}


\maketitle


\section{Sample structure and micro-magnet simulation}

The layer sequence of the device is shown in Fig. S1A.
The undoped Si/SiGe heterostructure used in this study is grown by chemical vapour deposition. 
First a 3 $\mathrm{\mu m}$ thick graded buffer is grown on a highly-resistive Si substrate
by linearly increasing the Ge content from 0 \% to 30 \%, and then a 1 $\mathrm{\mu m}$ thick Si$_{0.7}$Ge$_{0.3}$ buffer is grown. 
Next, a 15 nm thick Si quantum well, a 60 nm thick undoped Si$_{0.7}$Ge$_{0.3}$ spacer, and a 2 nm thick Si cap are successively grown on the buffer.
The resulting two-dimensional electron gas has a typical electron density of 5.0 $\times$ 10$^{11}$  cm$^{-2}$ and a mobility of 1.8 $\times$ 10$^{5}$ cm$^2$/Vs at an accumulation gate voltage of 1.5 V and a bath temperature of  $T=2.3$ K.

The surface of the heterostructure is covered by a 10 nm thick Al$_2$O$_3$ insulator formed by atomic layer deposition.
The ohmic contacts are fabricated by Phosphorous ion implantation.
The quantum dot confinement gates and the accumulation gate are formed by electron-beam lithography and metal deposition.
The accumulation gate and depletion gate electrodes are separated from each other by another 50 nm thick Al$_2$O$_3$ insulator layer.
A 250 nm thick cobalt micromagnet is deposited on top of the accumulation gate to induce a stray  magnetic field around the quantum dot.
The distance between the micromagnet and the Si quantum well is 162 nm.

The micromagnet design is adopted from our previous GaAs double quantum dot device \cite{Yoneda2014} (Fig. S1B).
The design is optimized to obtain the large slanting field and the local Zeeman field difference to maximize spin rotation speed and to minimize the single-qubit operation crosstalk errors.
The micromagnet magnetic field simulation is performed by using the method \cite{Radia} as in Refs. \cite{Yoneda2014, Yoneda2015}.
According to the calculation, the slanting magnetic field is 
$\mathrm{d}B^{\mathrm{MM,R}} _{\mathrm{y}}/\mathrm{d}z=0.8$ T/$\mu$m for the right quantum dot and $\mathrm{d}B^{\mathrm{MM,L}} _{\mathrm{y}}/\mathrm{d}z=0.75$ T/$\mu$m for the left quantum dot and the local Zeeman field difference is $| B^{\mathrm{MM,R}} _{\mathrm{z}} -B^{\mathrm{MM,L}} _{\mathrm{z}} |= 29$ mT.
The local Zeeman field difference is roughly consistent with our observation (see Fig. 1D in the main text).
On the other hand, estimating the slanting field from the experimental data is difficult, since $f_{\mathrm{Rabi}}$ is affected by many parameters such as the microwave gate lever arm for the dot position, the electric field direction at the dot position etc., which are difficult to measure experimentally.

The valley splitting is confirmed to be larger than the Zeeman splitting by magneto-spectroscopy measurement.
Therefore the physics in this work is mainly described by a conventional single-valley picture, although there may be a small fraction of the population in the excited valley state due to the initialization error.

\section{Measurement setup}

The sample is cooled down using a dilution refrigerator to a base electron temperature of 120 mK which is estimated from the transport line-width.
The gates R, L and C are connected to high-frequency coaxial lines for application of the gate voltage pulse and the microwave burst.
The high-frequency lines are attenuated inside the dilution refrigerator to dissipate the Johnson-Nyquist and technical noises from the room-temperature electronics ($-33$ dB for gates R and L, $-13$ dB for gate C).
The voltage pulse to the gate electrodes is generated by a Sony/Tektronix AWG520 arbitrary waveform generator, and applied via cryogenic bias-tees with a cut-off frequency of $\sim 50$ Hz.
The microwave signal is applied to the gate C using an Agilent E8267D vector microwave signal generator.
The output power is fixed at +16 dBm except for the randomized benchmarking measurement which requires fine tuning of the microwave amplitude $A_{\mathrm{MW}}$ to keep $T_{\pi} = 50$ ns.
The baseband signal for I/Q modulation is generated by another Sony/Tektronix AWG520 which is synchronized with the one  used for the gate voltage pulse generation.
The microwave signal is single side-band modulated from the local oscillator frequency by applying 80 MHz cosine/sine waveforms to the I/Q ports of the vector signal generator. The unwanted spurious signals are suppressed by applying proper frequency dependent corrections for DC offset, quadrature amplitude and phase.

Rapid measurement of the charge state is performed by rf-reflectometry of a sensor quantum dot coupled to an impedance matched resonant circuit which consists of a commercial 1.2 $\mu$H SMD inductor and a parasitic capacitance. 
The resonant circuit operates at its resonance frequency of 206.7 MHz.
The reflected signal is amplified by a cryogenic amplifier (Caltech CITLF1) mounted at the 4 K stage of the dilution refrigerator and further amplified and demodulated by room temperature electronics.
After the filtering (cutoff frequency of 300 (30) kHz for the left (right) dot), the signal is digitized using an Alazartech ATS9440 digitizer at a sampling rate of 5 MS/s.

Figure S2 shows a typical charge sensor response for the single-shot spin detection.
When a spin up state is measured, the charge sensor signal first increases as it tunnels out to reservoir before decreasing to its initial value when a down spin tunnels back into the dot.
To obtain the up spin probability $P_{\uparrow } ^{\mathrm{R(L)}}$, the measurement sequence is repeated for 250 to 1000 times per one data point.

\section{Right dot measurement data}

Figure S3A shows the measurement result of Rabi oscillations for the right quantum dot.
Since the right dot is far from the charge sensor quantum dot placed at the left side of the device, the sensitivity of the sensor is smaller than for the left dot. 
This causes the reduced readout visibility since a reduced measurement measurement bandwidth (30 kHz) has to be used to keep large enough signal to noise ratio. In this case, some of the fast tunneling events are not measured.
The visibility could be enhanced by retuning the gate voltages to reduce the right dot tunnel rate slower than the measurement bandwidth.

Figure S3B shows the Ramsey measurement data. The fit shows a Gaussian decay curve with an exponent of 2.
The measured $T_2 ^*$ of 2.2 $\mu$s is comparable to that of the left dot electron spin.

\section{Discussions on the microwave power dependence}

Dephasing of an electron spin in nuclear spin bath is usually dominated by slow fluctuation of nuclear spin  magnetic field.
Under the Rabi driving field $hf_{\mathrm{Rabi}}$, the Hamiltonian in the rotating frame of $f_{\mathrm{MW}}$ can be written as follows,
\begin{equation}
H=(hf_{\mathrm{Rabi}}) \sigma _{\mathrm{x}} + (h\Delta f) \sigma _{\mathrm{z}} ,
\end{equation}
where $\Delta f = f_{\mathrm{MW}}- f_{\mathrm{res}}$ is frequency detuning.
Although the nuclear spin bath can fluctuate $\Delta f$, the effect of the fluctuation can be strongly suppressed by the Rabi driving term $hf_{\mathrm{Rabi}} \sigma _{\mathrm{x}}$.

From the Ramsey data of the left quantum dot, the standard deviation of the nuclear field $\sigma = 1/ (\sqrt{2} \pi T_2 ^* ) = 122$ kHz is derived. This is more than an order magnitude smaller as compared to the typical Rabi driving used for our measurements ($>$ MHz).
The simulation of the nuclear spin effect for a Rabi driving with $f_{\mathrm{Rabi}}= 9$ MHz is performed 
using the following equation,
\begin{equation}
P_{\uparrow} (t_{\mathrm{p}}) = A + B \sum _{\Delta f=- 5\sigma} ^{ 5\sigma}
g_{\sigma} (\Delta f) \frac{(f_{\mathrm{Rabi}}) ^2}{(f_{\mathrm{Rabi}}) ^2 + (\Delta f) ^2}  
\mathrm{sin} (2\pi \sqrt{(f_{\mathrm{Rabi}}) ^2 + (\Delta f) ^2} t_{\mathrm{p}}), 
\end{equation}
where $g_{\sigma}(\Delta f)$ is a Gaussian distribution with standard deviation of $\sigma$ and $A$, $B$ are constants that account for the readout and initialization fidelities.
A step of $\delta (\Delta f) = 10 \sigma /2000$ is used for the summation to average 2,001 nuclear spin distributions.
The simulation result is shown in Fig. S3a.
As expected, since $\sigma$ is much smaller than the Rabi driving, no notable decay is observed in the simulated curve, although a clear decay is observed in the measurement data.
To fit the data, although the underlying mechanism of the decay is unknown, we empirically find that an exponential decay with an exponent 1 fits better than the Gaussian decay with an exponent of 2.
The fitting to the experimental data is performed using the following procedure,\\

1. We fit the raw data with exponential function to obtain initial guesses of $T_2 ^{\mathrm{Rabi}}$ and $f_{\mathrm{Rabi}}$.
\begin{equation}
\label{fitcurve}
P_{\uparrow} (t_{\mathrm{p}})= A \exp{\biggl( -\frac{t_{\mathrm{p}}}{T_2 ^{\mathrm{Rabi}}} \biggr)} \sin{(2\pi f_{\mathrm{Rabi}} t_{\mathrm{p}}+\phi )}+ B.
\end{equation}

2. We apply a finite impulse response (FIR) high-pass filter with a cutoff frequency of $f_{\mathrm{Rabi}}/2$ and a passband frequency of $f_{\mathrm{Rabi}}$ to the raw experimental data for $A_\mathrm{\mathrm{MW}} \geq 0.3$ in order to remove the gradual background change. 

3. We fit the FIR filtered data with an exponential function to obtain experimental $T_2 ^{\mathrm{Rabi}}$ and $f_{\mathrm{Rabi}}$\\

The use of the FIR filter is essential to obtain good fitting curves. Note that it does not change the damped oscillation term in the unfiltered raw data since the spectrum of the damped oscillations is centred at $f=f_{\mathrm{Rabi}}$ and has a typical width of $1/T_2 ^{\mathrm{Rabi}} < f_{\mathrm{Rabi}}$. 
The gradual background change is presumably due to the small electron temperature increase during the readout stage.

From the fitting of the data in Fig. 3C in the main text, we notice the empirical dependence of 
$T_2 ^{\mathrm{Rabi}} \propto (A_{\mathrm{MW}} )^{-4}$ (Fig. S3C).
If the system temperature $\mathcal{T}$ is determined by the microwave heating, the temperature should be proportional to the microwave power $(A_{\mathrm{MW}}) ^2$. 
Then the relation $T_2 ^{\mathrm{Rabi}} \propto \mathcal{T} ^{-2}$ between the Rabi decay time and the temperature can be obtained.
This is consistent with the observation by Dial {\it et al.} \cite{Dial2013} where the relation $T_2 ^{\mathrm{echo}} \propto \mathcal{T} ^{-2}$ is observed for a singlet-triplet qubit.

As mentioned in the main text, the photon-assisted tunneling can be another dephasing mechanism.
In a double quantum dot connected to reservoirs, there are two types of photon-assisted processes.
One is the inter-dot photon-assisted tunneling.
Since this tunneling process occurs only when the inter-dot energy detuning is the microwave frequency, this process can easily be ruled out by the outcome of the spectroscopy measurement (Fig. 1D in the main text).
Indeed, such a strong frequency dependent signal is not observed in the measurement data.
The other photon-assisted tunneling causes the electron exchange between the dot and the reservoir.
To exclude this possibility, the measurement to confirm the effect of the coupling to the reservoir is performed (Fig. S5).
The Rabi decay time measurement is performed for two different operation points with two different Coulomb blockade depths (about 400 or 200 $\mu$eV away from the (1,1)-(0,1) charge transition).
Due to the slight quantum dot displacement due to the operation point change, the centre resonance frequency of EDSR  shifts accordingly.
Since the microwave transmission is slightly different for the two different centre resonance frequencies, $T_2 ^{\mathrm{Rabi}}$ is plotted as a function of $f_{\mathrm{Rabi}}$ to account for this change.
The measurement result shows basically the same characteristics for the two different operation points, therefore it is unlikely the coupling to the reservoir causes the Rabi decay time 
decrease measured in this work.

\section{Randomized benchmarking}

Each data point in Fig. 4 of the main text is obtained from 12,800 single-shot measurements (800 points per one sequence).
The last recovery Clifford gate ensures that the ideal final state to be an eigenstate of $\sigma _{\mathrm{z}}$ which is up or down spin state.
In this measurement, the gates I, $\pm$X, $\pm$Y, $\pm$X/2 and $\pm$Y/2 are used as primitive Clifford gates since these can be implemented by single-step Gaussian microwave bursts.
The decomposition of the total 24 single-qubit Clifford gates by these primitive gates results in the average number of primitive gates per one decomposed Clifford gate of 1.875 \cite{Muhonen2015, Magesan2012}.
The single Clifford gate fidelity $F_{\mathrm{c}} ^{\mathrm{single}}$ refers to the average gate fidelity per one primitive Clifford gate while the Clifford gate fidelity $F_{\mathrm{c}}$ refers to the average gate fidelity per one decomposed Clifford gate.
From the measurement data of the reference sequence, the exponential decay curve $F(m) = A(p_{\mathrm{c}}) ^m$ and  the Clifford gate fidelity $F_{\mathrm{c}}=(1+p_{\mathrm{c}})/2$ can be obtained.
Then the single Clifford gate fidelity $F_{\mathrm{c}} ^{\mathrm{single}}$ is calculated using the following equation,
\begin{equation}
F_{\mathrm{c}} ^{\mathrm{single}}=\frac{1+p_{\mathrm{c}}^{\mathrm{single}}}{2} \sim 1-\frac{1-F_{\mathrm{c}}}{1.875},
\end{equation}
where $p_{\mathrm{c}}^{\mathrm{single}}= (p_{\mathrm{c}})^{\frac{1}{1.875}}$ represents the decrease of the sequence fidelity per single primitive Clifford gate.

To characterize each of the primitive Clifford gate fidelities, the interleaved randomized benchmarking \cite{Magesan2012a} is used.
In this measurement, similar exponential decay curves $F(m) = A(p_{\mathrm{gate}}) ^m$ are obtained and the fidelities for each interleaved gate $F_{\mathrm{gate}}$ are calculated from the following equation, 
\begin{equation}
F_{\mathrm{gate}} =\frac{1+(p_{\mathrm{gate}}/p_{\mathrm{c}})}{2}, 
\end{equation}
where the effect of the randomizing sequence is taken into account by the factor of $1/p_{\mathrm{c}}$.

\bibliography{library}

\newpage

\begin{figure}[t]
\centering
\includegraphics*[width=0.5\columnwidth]{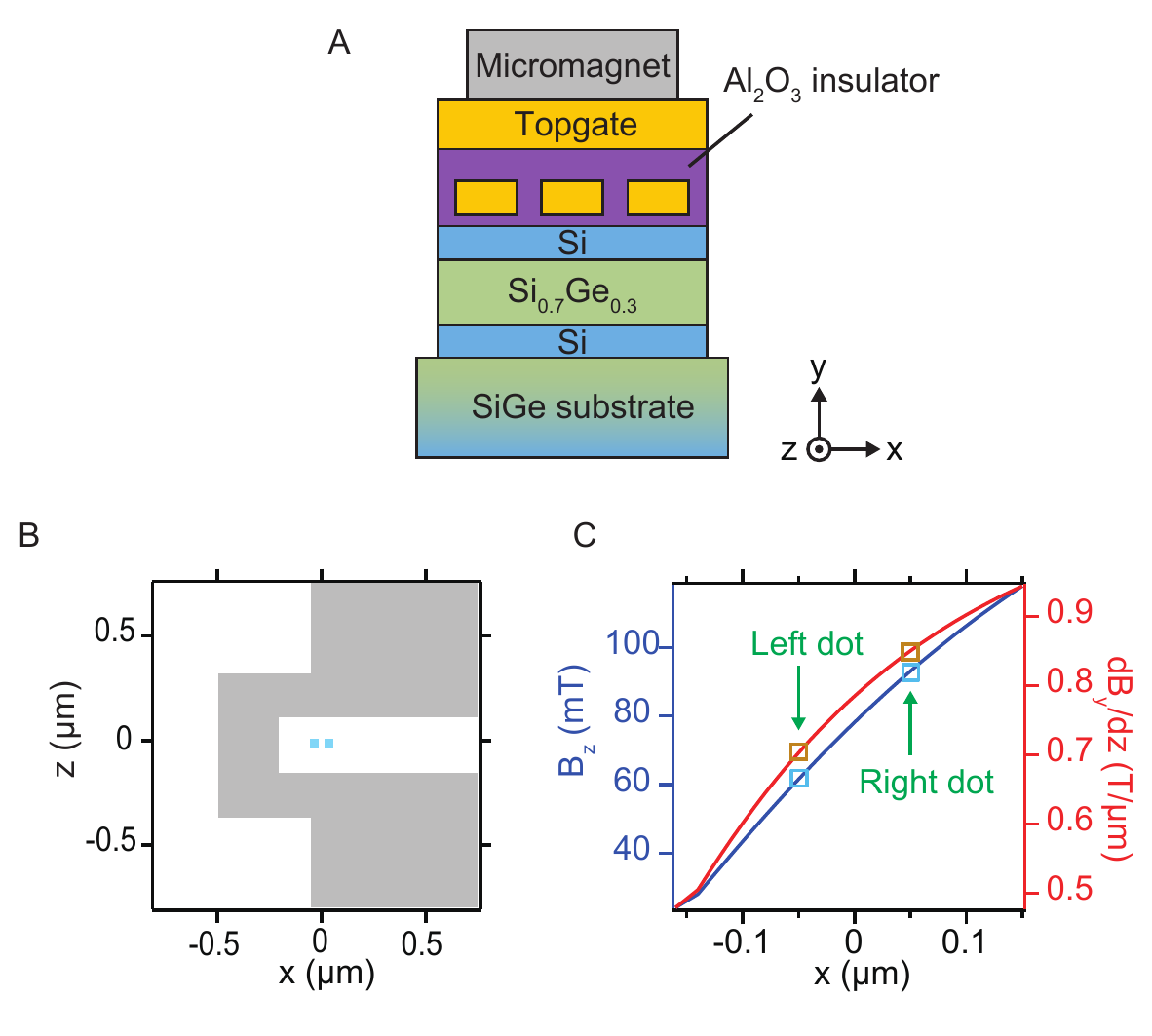}
\caption{\label{FigS1}
{\bf Micromagnet design and simulation}. 
{\bf A}, Schematic layer sequence of the device structure. The external magnetic field is applied along the positive z-direction.
{\bf B}, Schematic of the micromagnet design. The grey area shows the micromagnet pattern and the quantum dot locations are represented by the blue boxes separated by 0.1 $\mu$m.
{\bf C}, Simulated slanting magnetic field $\mathrm{d}B^{\mathrm{MM,R}} _{\mathrm{y}}/\mathrm{d}z$ (red curve) and  local Zeeman  field $B_{\mathrm{z}} ^{\mathrm{MM}}$ (blue curve) as a function of lateral the dot position $x$.
The left dot position is $x=-0.05$ $\mu$m and the right dot position is $x=0.05$ $\mu$m.}
\end{figure}

\begin{figure}[t]
\centering
\includegraphics*[width=0.5\columnwidth]{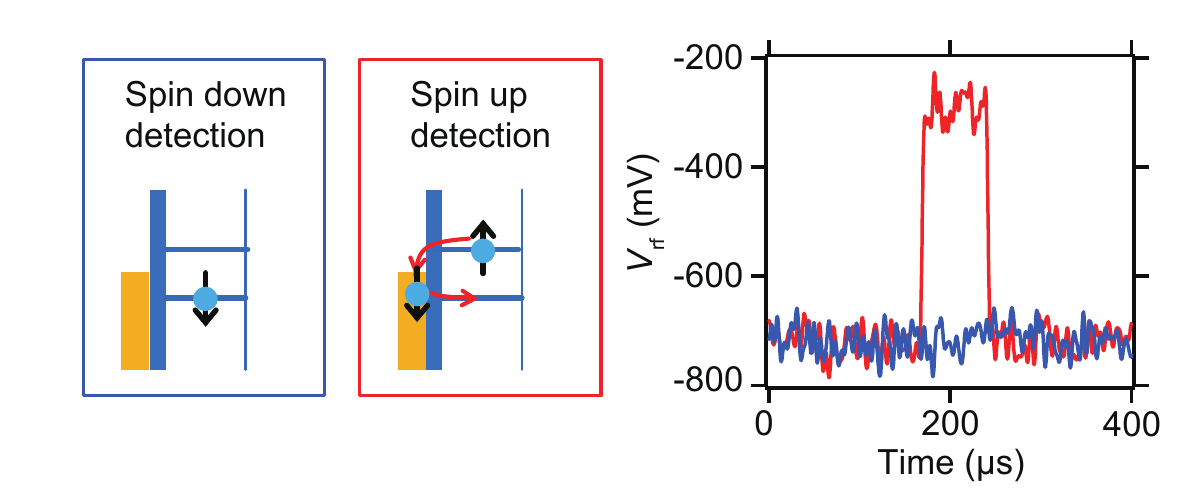}
\caption{\label{FigS2}
{\bf Single-shot spin readout using energy selective readout technique}. 
The pulse sequence in Fig 2A without the microwave burst is used for this measurement.
The blue trace corresponds to the case of spin-down state readout and the red trace corresponds to the spin-up state.
When a spin-up state is measured, the charge sensor signal first increases as the electron tunnels out to reservoir before decreasing to its initial value when another spin-down electron tunnels back into the dot.}
\end{figure}

\begin{figure}[t]
\centering
\includegraphics*[width=0.5\columnwidth]{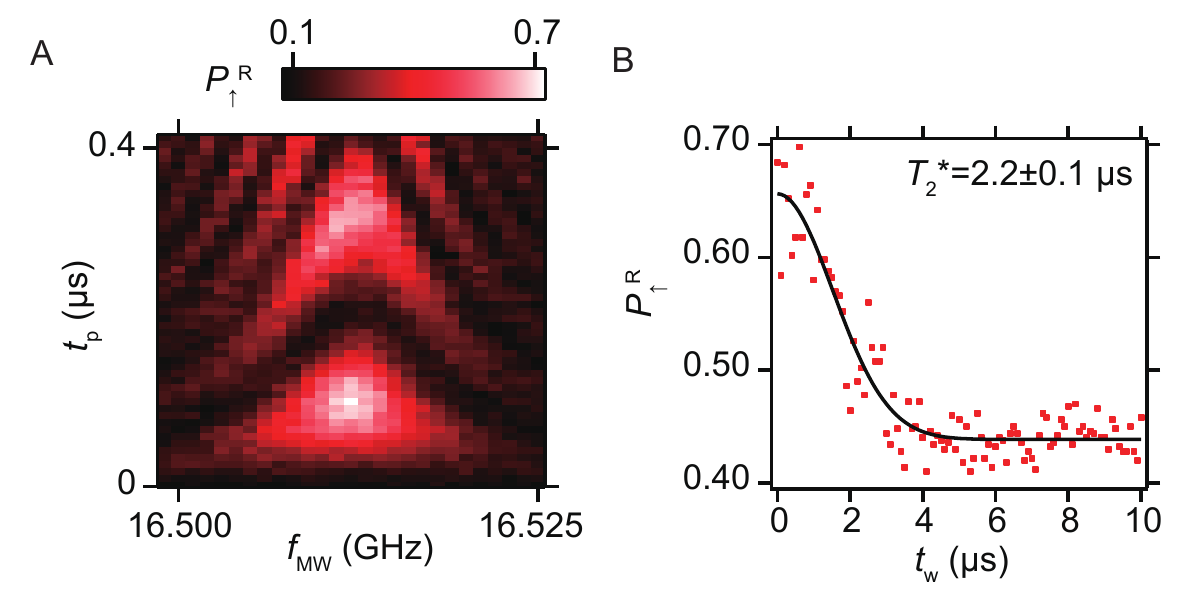}
\caption{\label{FigS3}
{\bf Rabi oscillation and Ramsey measurements of the right quantum dot}. 
{\bf A}, Measurement result of Rabi oscillation of the right dot spin state measured at $B_{\mathrm{ext}} = 0.512$ T. 
{\bf B}, Ramsey measurement result of the right dot spin qubit. The red points show the measurement data and the black solid line shows a Gaussian fitting curve. The obtained 2.2 $\mu$s phase coherence time is comparable to that (1.84 $\mu$s) in the left quantum dot.
}
\end{figure}

\begin{figure}[t]
\centering
\includegraphics*[width=0.5\columnwidth]{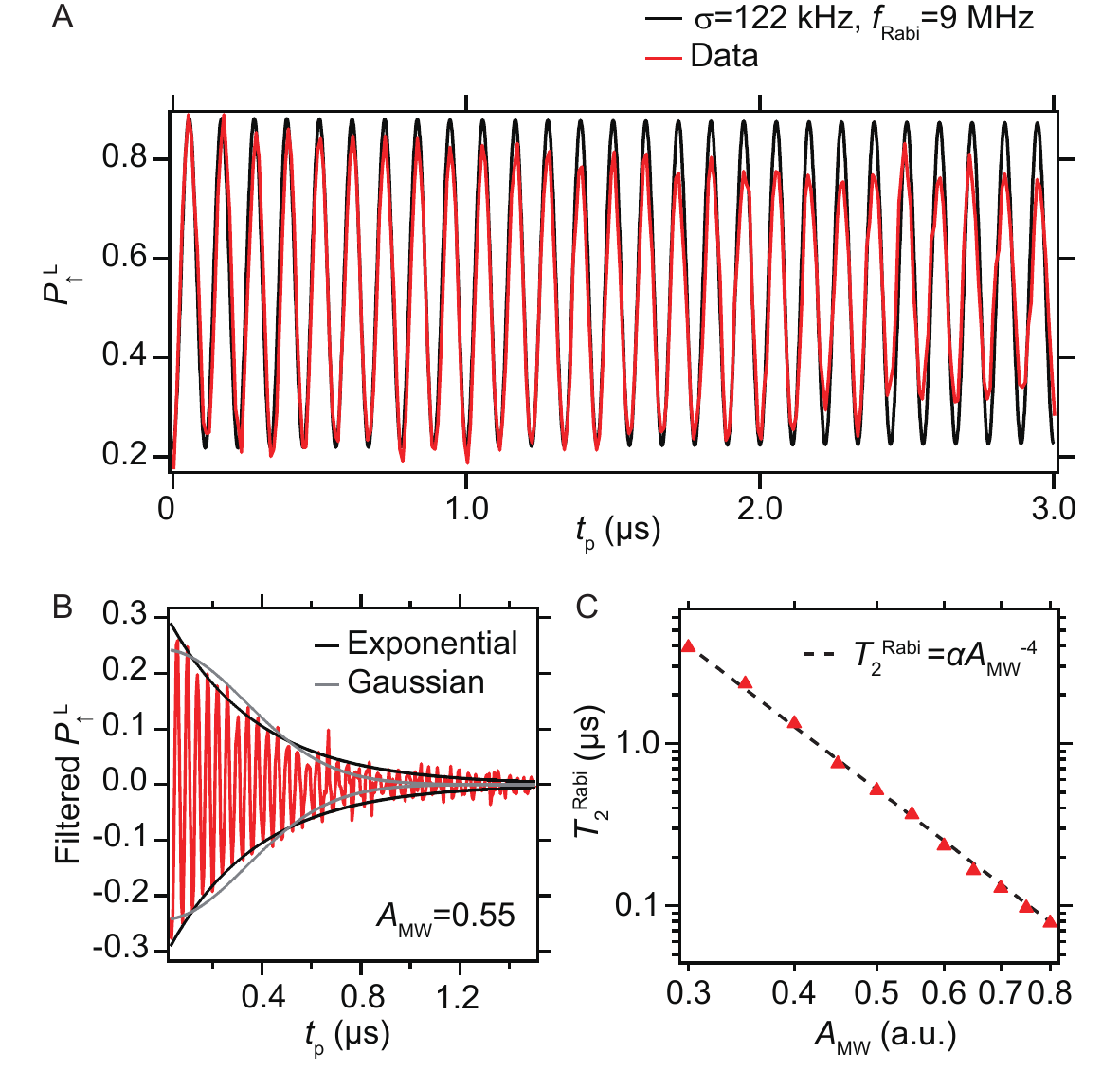}
\caption{\label{FigS4} 
{\bf Fitting of Rabi oscillation data}. 
{\bf A}, Comparison of the measured Rabi oscillation and the model calculation with a Gaussian nuclear field fluctuation.
The red solid line shows a experimental data with $f_{\mathrm{Rabi}}=9$ MHz (the data in Fig. 2c in the main text). The black solid line is a numerical calculation result using the parameters $\sigma = 122$ kHz and $f_{\mathrm{Rabi}}=9$ MHz.
{\bf B}, Fitting of the damped oscillation at $A_{\mathrm{MW}} = 0.55$. The black solid line shows a exponential envelope $\pm A \exp{(-t_{\mathrm{p}} / T_2 ^{\mathrm{Rabi}})} + B$ and the grey solid line shows a Gaussian envelope $\pm A \exp{(-(t_{\mathrm{p}} / T_2 ^{\mathrm{Rabi}} )^2)} + B$.
{\bf C}, Microwave amplitude dependence of the Rabi oscillation decay time. The red triangles show the data and the black dotted line shows fitting with a function $ T_2 ^{\mathrm{Rabi}} = \alpha (A_{\mathrm{MW}}) ^{-4}$. 
}
\end{figure}

\begin{figure}[t]
\centering
\includegraphics*[width=0.5\columnwidth]{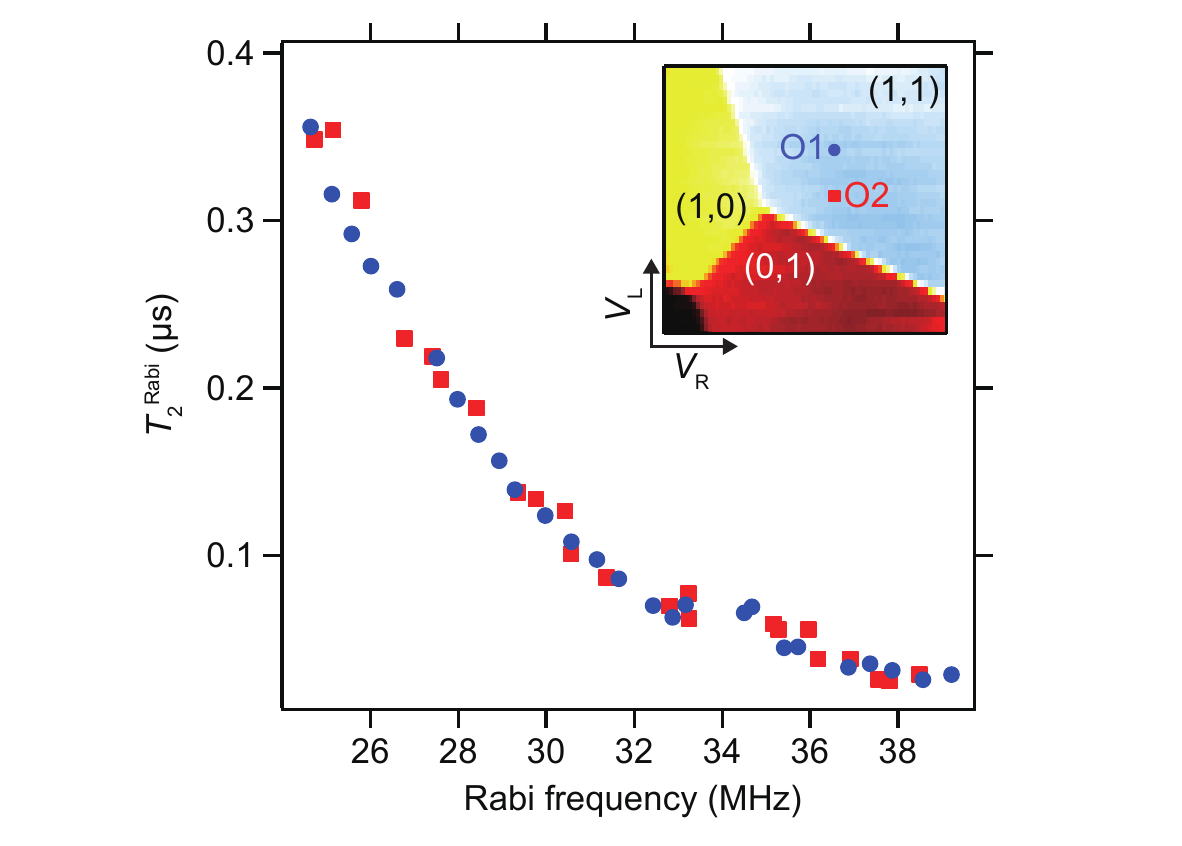}
\caption{\label{FigS5}
{\bf Rabi decay measurement for two different operation points}. 
The operation point 1(2) (O1(O2) in the inset charge stability diagram) is about 400(200) $\mu$eV away from the (1,1)-(0,1) charge transition line.
The two data sets basically showing almost the same characteristic, therefore the effect of the coupling to the reservoir is small for the enhanced dephasing at high microwave amplitudes. } 
\end{figure}